\documentstyle[12pt,aaspp4,psfig]{article} 

\received{3 November 2000} 
\accepted{27 December 2000} 
 
\slugcomment{Astronomical Journal} 
 
\begin{document} 

\title{Thermal Infrared 3--5 $\mu$m Colors of Obscured and Unobscured 
Active Galactic Nuclei}
\author{Masatoshi Imanishi \altaffilmark{1,2}}  
\affil{National Astronomical Observatory, Mitaka, Tokyo 181-8588, Japan}
 
\altaffiltext{1}{Institute for Astronomy, University of Hawaii, 
2680 Woodlawn Drive, Honolulu, Hawaii 96822, USA}
\altaffiltext{2}{Visiting Astronomer at the Infrared Telescope Facility 
which is operated by the University of Hawaii under contract to the 
National Aeronautics and Space Administration.}

\begin{abstract} 
 
Thermal infrared photometry in the $L$- and $M'$-band and $L - M'$ 
colors of type-1 and type-2 active galactic nuclei (AGNs) are 
presented.
After combining our observations with photometric data at similar 
wavelengths taken from the literature, 
we find that the excess of $L - M'$ colors of type-2 AGNs 
(37 sources, 50 data points) relative to type-1 AGNs 
(27 sources, 36 data points), due to dust extinction, is statistically 
detectable, but very small.
We next investigate the $L - M'$ colors of type-2 AGNs 
by separating less dust-obscured type-2 AGNs and highly dust-obscured 
type-2 AGNs.  
In both cases, the $L - M'$ colors are similar to the intrinsic 
$L - M'$ color of unobscured AGNs, and 
the $L - M'$ color excess of the latter highly dust-obscured type-2 AGNs 
due to dust extinction is much smaller than that expected from 
the Galactic dust extinction curve.
Contamination from starbursts and the time lag of flux variation 
are unlikely to explain this small $L - M'$ color excess, 
which is best explained if the dust extinction curve in the close 
vicinity of AGNs is fairly flat at 3--5 $\mu$m as a result of a size 
increase of the absorbing dust grains through coagulation.

\end{abstract} 
 
 
\section{Introduction} 

According to the current unification paradigm for active galactic
nuclei (AGNs), type-1 AGNs (which show broad optical emission lines)
and type-2 AGNs (which do not) are intrinsically the same, but the
nuclei of the latter class are obscured by dust that lies along our
line of sight in dusty molecular tori close to the AGNs (\cite{ant93}).  
Estimation of the amount of dust along our line of sight in
type-2 AGNs and comparison with the amount in type-1 AGNs is an
important observational test of the unification paradigm.  A direct
estimate of dust extinction toward highly luminous type-2 AGNs is
necessary to answer the question ``how common are highly luminous and
highly dust-obscured AGNs (so-called type-2 quasars)?'' (\cite{hal99}).

X-ray spectroscopic observations of type-2 AGNs imply higher X-ray
absorption columns than do observations of type-1 AGNs, supporting the
unification paradigm (e.g., \cite{nan94,smi96}).  
However, X-ray absorption is caused both by dust and gas.
Estimating the amount of dust along our line of sight ($A_{\rm V}$) 
from X-ray absorption ($N_{\rm H}$) is uncertain, since the
$N_{\rm H}$/$A_{\rm V}$ ratios toward AGNs are found to vary by more
than an order of magnitude (\cite{alo97}).

For several reasons, we would expect study in the thermal infrared
(3--5 $\mu$m) wavelength range to be a powerful tool for estimation 
of dust extinction toward AGNs.  
Firstly, flux attenuation in this band is
purely caused by dust extinction, and the effects of dust extinction
are wavelength dependent in the Galactic diffuse interstellar medium
(\cite{rie85,lut96}). Secondly, the
absolute flux attenuation by dust extinction is smaller than at
shorter wavelengths (\cite{rie85,lut96}).
Thirdly, extended stellar emission generally dominates over obscured
AGN emission at $<$2 $\mu$m, whereas at $>$3 $\mu$m moderately
luminous obscured AGNs show compact, AGN-related emission that
dominates observed fluxes (Alonso-Herrero et al. 1998, 2000, 
\cite{sim98a}, but see Simpson, Ward, \& Wall 2000).
Finally, since the compact emission at 3--5 $\mu$m most likely 
originates in hot 
(600--1000 K) dust at a part of the dusty molecular torus very near to
the AGN (close to the innermost dust sublimation region), the dust
extinction toward the 3--5 $\mu$m emission region is almost the same as
that toward the central engine itself.  
Hence, by comparing observed continuum fluxes at more than one 
wavelength between 3 and 5 $\mu$m, 
we can estimate dust extinction toward obscured AGNs directly, up to
high magnitudes of obscuration, without serious uncertainties in the
subtraction of stellar emission.  Some attempts to estimate dust
extinction toward obscured AGNs have been made based on near-infrared 
1--5 $\mu$m colors (\cite{sim98a,sim99,sim00}), but, given that 
only upper limits are available at 3--5
$\mu$m in most cases, the estimate depends heavily on data at $<$3
$\mu$m in the rest-frame, where stellar emission dominates the
observed fluxes.

We have conducted $L$ (3.5$\pm$0.3 $\mu$m) and $M'$ (4.7$\pm$0.1 $\mu$m) 
band photometry of type-1 and type-2 AGNs.
The main aim is to investigate the $L - M'$ colors of a large number of 
type-1 and type-2 AGNs and to examine the question of whether 
$L - M'$ colors are a good measure of the dust extinction toward AGNs.
Throughout this paper, $H_{0}$ $=$ 75 km s$^{-1}$ Mpc$^{-1}$,  
$\Omega_{\rm M}$ = 0.3, and $\Omega_{\rm \lambda}$ = 0.7 are adopted. 

\section{Target Selection}

The target sources were selected based on their proximity and 
high optical [OIII] emission line luminosities.
The first and second criteria were adopted, respectively, to make 
detection at $M'$ feasible and to select reasonably luminous 
AGNs (\cite{sim98b}), for which contamination from extended 
star-formation-related emission (both stellar emission and dust emission 
powered by star-formation activity) is expected to be 
smaller than it would be for less luminous AGNs.
Our samples are heterogeneous and not statistically complete, but 
provide useful information on the $L - M'$ colors of AGNs.

\section{Observation and Data Analysis}

$L$ (3.5$\pm$0.3 $\mu$m) and $M'$ (4.7$\pm$0.1 $\mu$m) band  
photometry was performed at the NASA Infrared Telescope Facility (IRTF) 
using NSFCAM (\cite{shu94}).
Table 1 gives details of the observations.
Sky conditions were photometric throughout the observing runs.
The seeing sizes measured from standard stars were 
0$\farcs$6--1$\farcs$2. 
The NSFCAM used a 256$\times$256 InSb array.
For $M'$-band photometry, the smallest pixel scale 
(0$\farcs$06 pix$^{-1}$) was used during all the observing runs.
For $L$-band photometry, the pixel scale of 0$\farcs$06 pix$^{-1}$ was 
used in November 1999, while that of 0$\farcs$15 pix$^{-1}$ was used 
in April and May 2000.
The field of view is 14$''$ $\times$ 14$''$ and 
38$''$ $\times$ 38$''$ in the case of  0$\farcs$06 pix$^{-1}$ and 
0$\farcs$15 pix$^{-1}$ pixel scales, respectively.
Each exposure was 0.3--0.4 sec long at $L$ and 0.12--0.2 sec at $M'$.
A dithering technique was utilized with an amplitude of 
3--10$''$ to place sources at five different positions on the array.
At each dithering position, 50--200 frames were coadded.
Offset guide stars were used whenever available to achieve high 
telescope tracking accuracy.

Standard data analysis procedures were employed, using IRAF
%-------------
\footnote{
IRAF is distributed by the National Optical Astronomy Observatories, 
which are operated by the Association of Universities for Research 
in Astronomy, Inc (AURA), under a cooperative agreement with the 
National Science Foundation. 
}.  
%-------------
Firstly, bad pixels were removed and the values of these pixels
were replaced with interpolated values from the surrounding pixels.
Secondly, the frames were dark-subtracted and then scaled to have the
same median pixel value, so as to produce a flat frame.  
The dark-subtracted frames were then divided by a normalized flat frame 
to produce images at each dithering position.  Standard stars and very
bright AGNs were clearly seen in images at each dithering position,
and so images that contained these were aligned to sub-pixel accuracy
using these detected sources and then summed to produce the final
images.  However, for fainter AGNs, the sources were not always
clearly recognizable in the individual images at each dithering
position. In these cases the images were shifted based on the records
of telescope offset, assuming that telescope pointing and tracking
were accurate, and were then summed to produce final images.  This
procedure potentially broadens the effective point spread function in
the final image, providing larger source full widths at half maximum
(FWHMs) than the expected values.

At 3--5 $\mu$m, and particularly at $M'$, 
thermal emission from a small amount of occasionally 
transiting cirrus can increase sky background signals and affect 
data quality, even though the sky may look clear.
Thus, before summing the frames, we confirmed that their sky background 
levels agreed to within 1\%, showing that the data were not
seriously affected by this kind of cirrus.

The images of all the observed AGNs were spatially compact, 
with no clear extended emission found at either band.
The measured FWHMs of some AGNs in the final images were slightly larger 
than the FWHMs of standard stars, but
we attribute these larger FWHMs mainly to the uncertainty 
introduced by shifting and adding frames containing faint sources, as
discussed above.
Photometry was done with 6$''$ diameter apertures, by using 
the task ``PHOT''.
Since the 3--5 $\mu$m emission was compact, the resulting photometric 
magnitudes were almost independent of aperture size as long as the 
aperture sizes were sufficiently larger than the measured FWHMs.
Flux calibrations were made by using standard stars in the faint Elias 
standard star catalog  
%-----------
\footnote{
http://irtf.ifa.hawaii.edu/online/IRTF/Catalogs/Elias\_standards
}
%-----------
or IRTF bright infrared standard star catalog  
%-----------
\footnote{
http://irtf.ifa.hawaii.edu/online/IRTF/Catalogs/bright\_standards
}.
%-----------
The $M'$-band magnitudes of standard stars were assumed to be the same 
as the $M$-band magnitudes listed in these catalogs, 
because the $M - M'$ color is expected to be virtually 0 mag 
unless a standard star is of a very late type.

$M'$-band photometry is inherently very difficult because of the large
background noise.  The photometric accuracy was strongly dependent on
the uncorrectable spatial fluctuation of background signals;  because
of this uncorrectable fluctuation, $M'$-band photometric accuracy was 
not necessarily better in objects with longer integration time.
Furthermore, since the smallest pixel scale (0$\farcs$06 pix$^{-1}$)
had to be utilized in our observations to avoid saturation, emission
from the compact AGN emission was spread over many pixels, which made 
the recognition of real detections even more difficult.  To avoid
spurious detections, we divided $M'$-band data into two or three
independent images and confirmed that the source positions agreed with
each other.

\section{Results}

Our new photometric measurements for type-1 and type-2 AGNs are
tabulated in Table 2.  We estimate $L - M'$ colors based on $L$- and
$M'$-band photometric data taken on the same night or two successive
nights.  Figure 1a shows the distribution of $L - M'$ colors of the
AGNs measured with our standard 6$''$ diameter apertures.  
Based on photometry of quasars (at $<$2.2 $\mu$m, 3.7 $\mu$m, 
and 10.1 $\mu$m), Neugebauer et al. (1987) found that the continuum 
spectral energy distribution can be approximated by a power law of 
the form $F_{\nu} \propto \nu^{1.4\pm0.3}$ at 3--5 $\mu$m, which implies 
that the intrinsic $L - M'$ color of quasars is 1.0$\pm$0.1.  
Since this color was derived using quasars (that is,
highly luminous unobscured AGNs), any contamination from 
star-formation-related emission is expected to be very small.  
We adopt this value for the intrinsic $L - M'$ color of unobscured AGNs.
The type-1 and type-2 AGNs in Figure 1a both show $L- M'$ colors 
similar to the intrinsic $L - M'$ color of unobscured AGNs.

To increase the sample size, we search the available literature
for 3--5 $\mu$m photometric data on galaxies.  Table 3 summarizes 
the results for sources whose magnitudes have been measured both
at $\sim$3.5 $\mu$m ($L$- or $L'$-band) and at $\sim$4.7 $\mu$m 
($M'$- or $M$-band).  
We exclude sources for which only upper limits were
given at one wavelength. When photometry has been performed with
several different aperture sizes by the same authors, we have
tabulated the photometric results based on the smallest aperture, in
order to minimize contamination from the extended star-formation-related 
emission of the host galaxies, except in the case of NGC 1068, 
for which photometry with a
0$\farcs$6 aperture was adopted (\cite{mar00}). 
The aperture sizes used are 0$\farcs$6--22$\farcs$5.  
The central wavelength and wavelength coverage of the filters used 
for photometry at these two bands differ slightly among authors, and
magnitude conversion formulas between slightly different filters are
not well established.  
We therefore assume that photometric magnitudes
in slightly different filters are the same.  Although this assumption
might introduce an uncertainty in the final photometric magnitudes of
0.1 mag or so, this would not affect our conclusions.  Figure 1b plots 
$L - M'$ colors of these AGNs and starburst/LINER galaxies taken 
from the literature.

\section{Discussion} 

\subsection{Comparison of $L - M'$ Colors}

After combining our data with data in the literature, 
we find that, for type-1 AGNs (27 sources, 36 data points), 
type-2 AGNs (37 sources, 50 data points), and
starbursts/LINERs (22 sources, 23 data points), 
the median (mean) $L - M'$ colors are 0.9 (0.8), 0.9 (1.0), 
and 0.4 (0.3) respectively.  
The median $L - M'$ colors of both type-1 and type-2 AGNs 
are within the range of the intrinsic $L - M'$ color of unobscured 
AGNs (1.0$\pm$0.1), but that of starbursts/LINERs is clearly smaller 
than those of AGNs.
Since the relative contribution of star-formation-related emission 
is larger in starbursts and LINERs than in AGNs, this implies that 
star-formation-related emission gives rise to bluer $L - M'$ colors 
than do AGNs. 

Figure 2 shows the cumulative probability distribution of 
$L - M'$ colors of type-1 and type-2 AGNs.
We apply the Kolmogorov-Smirnov test and find that the probability 
that the two distributions are drawn from different distributions 
is 88\%.
Thus, statistically, the $L - M'$ colors of type-2 AGNs are different 
from those of type-1 AGNs.
However, the difference appears to be very small, 
the median (mean) being 0.0 (0.2) mag redder in the former 
than the latter. 
If the Galactic dust extinction curve 
($A_{\rm L}$/$A_{\rm V}$ = 0.058, 
$A_{\rm M'}$/$A_{\rm V}$ = 0.023; \cite{rie85}) 
is adopted, the 0.2 mag difference in the $L - M'$ colors
implies that type-2 AGNs have higher dust extinction than 
type-1 AGNs with only $A_{\rm V}$ $\sim$ 6 mag, typically.

We next investigate the ``physical'' aperture sizes used for the 
measurements of the $L - M'$ colors of the AGNs, 
because, when larger physical aperture sizes are used, 
contamination from extended star-formation-related emission 
could be larger, which might decrease the $L - M'$ colors.
The median physical aperture sizes are 3.0 kpc and 1.3 kpc 
for type-1 and type-2 AGNs, respectively.
Figure 3 compares the cumulative probability distribution of the physical 
aperture size between type-1 and type-2 AGNs.
We do not find any clear trend that physical aperture size is 
systematically larger for type-2 AGNs than for type-1 AGNs. 
Furthermore, for type-1 AGNs, the median $L - M'$ colors below and above 
the median physical aperture size (3.0 kpc) are 0.7 mag and 0.9 mag, 
respectively.
For type-2 AGNs, those below and above the median physical aperture size  
(1.3 kpc) are 0.9 mag and 1.1 mag, respectively.
Therefore, $L - M'$ colors of AGNs measured with larger physical 
aperture sizes are not systematically bluer due to greater 
contamination from extended star-formation-related emission.
Thus, contamination from extended star-formation-related 
emission is unlikely to have strong, systematic effects on the 
$L - M'$ colors of AGNs.

We next investigate the $L - M'$ colors of type-2 AGNs, distinguishing
between those that are less dust-obscured and those highly dust-obscured,
because even though a galaxy may be classified as a type-2 AGN,
the dust extinction toward its nucleus could vary significantly.  
For some type-2 AGNs, dust extinction has been estimated to be high.  
IRAS 08572+3915, NGC 7172, and NGC 7479 display strong silicate dust 
absorption features at 9.7 $\mu$m (\cite{dud97,roc91}) and 
strong carbonaceous dust absorption at 3.4 $\mu$m (\cite{imd00,ima00}), 
which means that a large number of both carbonaceous and silicate 
dust grains lie in front of the background AGN emission. 
Since interstellar dust consists mainly
of carbonaceous and silicate dust grains (\cite{mat77,mat89}), 
the presence of many of these grains along
our line of sight to the AGN implies high dust extinction toward the
AGN emission.  
Besides the above sources, a clear 3.4 $\mu$m carbonaceous dust 
absorption feature 
is detected in the spectrum of NGC 1068 (\cite{bri94,ima97}).  
For Cygnus A, the observed $L$-band flux is significantly smaller than 
that expected from the intrinsic AGN power, which is estimated based on 
the optical [OIII] flux or extinction corrected 2--10 keV X-ray 
luminosity (\cite{war96}); 
it is argued that the small $L$-band flux is a result of flux
attenuation by dust extinction (\cite{war96}).  If the properties of the
obscuring dust are similar to those in the Galactic diffuse
interstellar medium ($\tau_{3.4}$/$A_{\rm V}$ $=$ 0.004--0.007;
\cite{pen94}, $\tau_{9.7}$/$A_{\rm V}$ = 0.05--0.1; 
Roche \& Aitken 1984, 1985, $A_{\rm L}$/$A_{\rm V}$ = 0.058; \cite{rie85})
%----------
\footnote{
$\tau_{3.4}$ and $\tau_{9.7}$ mean the optical depths of 
the 3.4 $\mu$m carbonaceous dust absorption and the 9.7 $\mu$m 
silicate dust absorption, respectively.
}, 
%----------
then the estimated column density of the obscuring dust is very large, 
corresponding to $A_{\rm V}$ = 140 mag for Cygnus A (Ward
1996), $A_{\rm V}$ $>$ 100 mag for IRAS 08572+3915 (\cite{imu00}), 
$A_{\rm V}$ = 30 mag for NGC 1068 (\cite{bri94,ima97,mar00}),
and $A_{\rm V}$ $>$ 20 mag for NGC 7172 and NGC 7479 
(\cite{roc91,ima00}).  
All of these sources are thus very likely to be 
highly dust-obscured AGNs.  

On the other hand, NGC 2992, 
IRAS 05189$-$2524, IRAS 20460+1925, MCG$-$5--23--16, and PKS 1345+12 
show detectable broad Pa$\alpha$ or Pa$\beta$ emission
lines at $<$2 $\mu$m 
(Veilleux, Goodrich, \& Hill 1997a, 
Veilleux, Sanders, \& Kim 1997b, 1999b).
These sources are thus classed as less dust-obscured type-2 AGNs.

In Figure 4 we plot the $L - M'$ colors of these representative samples 
of less dust-obscured and highly dust-obscured type-2 AGNs.
For both less dust-obscured and highly dust-obscured type-2 AGNs, 
the $L - M'$ colors are similar to the intrinsic $L - M'$ color of 
unobscured AGNs.
If the Galactic dust extinction curve of Rieke \& Lebofsky (1985) is 
applied, screen dust extinction with $A_{\rm V}$ = 50 mag should make 
the $L - M'$ color deviate from the intrinsic color with $\sim$1.7 mag.
The color deviation in the case of the Galactic dust extinction curve 
is so large that it should be easily recognizable in the highly 
dust-obscured type-2 AGNs in Figure 4.
The actual color deviation in the highly dust-obscured type-2 AGNs is, 
however, much smaller than that expected.

Dust obscuration toward type-2 AGNs might be due not only to the
dusty tori in the close vicinity of AGNs, but also to dust in the host
galaxies (on $>$100 pc scales).  In the latter case, a screen dust
extinction model is applicable.  Among our five highly dust-obscured
AGNs (IRAS 08572+3915, Cygnus A, NGC 1068, NGC 7172, and NGC 7479), 
NGC 7172 is thought to belong to this class of
object (\cite{ima00}).  In the former case, where obscuration comes
from the torus, the dust has a temperature gradient, with the inner
dust having a higher temperature (\cite{pie92}).  The $L$- and
$M'$-band emission is dominated by $\sim$900 K and $\sim$600 K dust,
respectively, and since the $M'$-band emitting dust is located further
out than the $L$-band emitting dust, $M'$-band emission suffers less
flux attenuation by dust extinction than $L$-band emission in type-2
AGNs.  
For IRAS 08572+3915, Cygnus A, NGC 1068, and NGC 7479, 
since dust extinction toward 3--4 $\mu$m emission region estimated 
using 3--4 $\mu$m data is larger than that toward $\sim$10 $\mu$m 
emission region estimated using $\sim$10 $\mu$m data, and/or that 
toward $\sim$10 $\mu$m emission region is larger than that 
toward $\sim$20 $\mu$m emission region estimated using 
$\sim$20 $\mu$m data, the presence of a temperature gradient in the 
obscuring dust is strongly suggested, indicating that these objects 
are obscured by dusty tori (\cite{dud97,imu00,ima00}).  
The presence of this temperature gradient in the obscuring dust should
increase (not decrease) the $L - M'$ colors compared to a screen dust
extinction model and thus cannot explain the small $L - M'$ colors
observed in these highly dust-obscured type-2 AGNs.

\subsection{Possible Reasons for the Small $L- M'$ Color Excess 
in Dust-Obscured AGNs}

\subsubsection{Time Lag between $L$- and $M'$-band Flux Variation}

According to the unification paradigm for AGNs, 3--5 $\mu$m emission
is dominated by thermal emission powered by UV to optical emission from
the central engine.  The UV to optical emission is known to be highly
time variable (e.g., \cite{cla92,nan98}).  Since
the $L$-band emission region ($\sim$900 K dust) is located closer to
the central engine than the $M'$-band emission region ($\sim$600 K
dust), a time lag in flux variation is expected, in the sense that the
$L$-band flux responds to the flux variation of the central UV to optical
emission prior to the $M'$-band flux.

A lower limit on the time lag is determined by the physical separation 
between the $L$- and $M'$-band emission regions.
The physical separation depends strongly on the UV to optical luminosity 
of the central engine and on the assumed dust radial density 
distribution in the dusty torus.
We use the code {\it DUSTY} (\cite{ive99}) to 
estimate the separation in the case of a reasonable dust spatial 
distribution. {\it DUSTY} solves the radiative transfer equation for a
source embedded in a spherically symmetric dusty envelope and 
calculates the resulting radial temperature distribution.

We use the same basic parameters for the spectral shape of the central
UV to optical emission and the dust composition as those adopted in
Imanishi \& Ueno (2000), and consider the UV to optical luminosities
of central engines with $>$ 10$^{11}L_{\odot}$.  We assume the ratio
of the outer to inner radius of the dusty envelope to be 200, where
the inner radius is determined by the UV to optical luminosity and by
the dust sublimation temperature, which is assumed to be $\sim$1000 K.
Using reasonable parameters, such as a dust extinction toward the
central engine of $A_{\rm V}$ = 0--200 mag and a radial dust density
distribution approximated by a power law ($\propto$ r$^{-\gamma}$)
with an index of $\gamma$ = 0--2, we find that the physical separation
between 900 K and 600 K dust is always larger than a few light days,
so the time lag is longer than a few days. The time lag of the
flux variation between the $L$- and $M'$-band could therefore affect the
$L - M'$ color measurements.  In fact, the $L - M'$ colors of sources
with multiple observations in Tables 2 and 3 are indeed different on
different observing dates, and these color differences could be
attributed to the time lag.  
If all the highly dust-obscured type-2 AGNs happened to be observed 
when they are bright at $L$ but faint at $M'$, 
then the derived $L - M'$ colors of these dust-obscured type-2 AGNs 
would be small. 
However, although this explanation cannot be ruled out completely,
it seems implausible.

\subsubsection{Contamination from Compact Nuclear Starbursts}

Compact 3--5 $\mu$m emission has hitherto been regarded as AGN-related
emission.  We now consider the possibility that this compact emission
may contain a significant contribution from compact nuclear starbursts.  
The presence of such compact nuclear starbursts has been suggested in 
some obscured AGNs (e.g., \cite{gon00}; but see \cite{iva00}).  

If galaxies possessed both AGN and (less obscured) nuclear starburst 
activity, and if the intrinsic magnitude ratios of these two components 
were the same among galaxies, then the contribution of AGN emission to 
observed 3--5 $\mu$m fluxes would be smaller in more highly 
dust-obscured AGNs 
as a result of the larger flux attenuation of AGN emission.  
Thus the $L - M'$ colors would not necessarily be larger in more highly 
dust-obscured AGNs.

When compact nuclear starbursts contribute significantly to
the 3--5 $\mu$m fluxes measured within the central few arcsec, the
nuclear 3--4 $\mu$m spectra are expected to display the 3.3 $\mu$m 
polycyclic aromatic hydrocarbon (PAH) emission (\cite{imd00}).  
However, the nuclear spectra of NGC 1068 (3$\farcs$8
$\times$ 3$\farcs$8), NGC 7172 (1$\farcs$2 $\times$ 5$''$), NGC 7479
(1$\farcs$2 $\times$ 5$''$), and IRAS 08572+3915 (1$\farcs$2 $\times$
8$''$) show clear 3.4 $\mu$m absorption feature but no detectable 3.3
$\mu$m PAH emission (\cite{ima97,ima00,imd00}), 
indicating that the observed 3--4 $\mu$m fluxes are
dominated by obscured AGN emission and not by starbursts.  
For Cygnus A, the emission at $>$3 $\mu$m is dominated by nuclear 
compact emission (\cite{djo91}, this work, \cite{imu00})
and the nuclear spectrum shows no detectable 11.3 $\mu$m PAH emission
(\cite{imu00}), suggesting that in this case too starburst
activity contributes little to the observed flux at $>$3 $\mu$m. 
We conclude that for these five highly dust-obscured AGNs (NGC 1068, 
NGC 7172, NGC 7479, IRAS 08572+3915, and Cygnus A), the small $L - M'$
colors are unlikely to be caused by the contamination from nuclear 
starburst activity.

\subsubsection{Flat Dust Extinction Curve at 3--5 $\mu$m}

All the above five sources except NGC 7172 are thought obscured
by dusty molecular tori in the close vicinity of AGNs (see \S 5.1).
High dust density and high turbulence velocity in these tori could
promote dust coagulation (\cite{ros91}), which might make the dust
size there much larger than that in the Galactic diffuse interstellar
medium (Maiolino et al. 2000a, b).  
If the size of a typical dust grain in
the AGNs' dusty tori is as large as a few $\mu$m, as suggested by
Maiolino et al. (2000a, b), the extinction curve at 3--5 $\mu$m could
be much flatter than that in the Galactic diffuse interstellar medium.
This flat dust extinction curve at 3--5 $\mu$m can explain the small 
$L - M'$ color excess in the highly dust-obscured AGNs. 
If this is the case, a strong caveat would need to be
attached to estimations of the dust extinction toward AGNs obscured by 
dusty tori that assume the applicability of a dust extinction curve
derived from the Galactic diffuse interstellar medium.

\section{Summary}  

$L - M'$ colors of obscured and unobscured AGNs were investigated. 
The $L - M'$ color excess in highly dust-obscured AGNs due to dust 
extinction, when compared to less dust-obscured AGNs,  
was much smaller than that expected from the Galactic dust extinction 
curve.
We argued that the size of the dust grains in the close vicinity of 
AGNs may be so large, due to coagulation, that the extinction curve 
at 3--5 $\mu$m is flatter than that in the Galactic diffuse 
interstellar medium.

\acknowledgments      

We thank P. Fukumura-Sawada, D. Griep and C. Kaminski for their 
support during the IRTF run.
We are grateful to Drs. J. Rayner and W. Vacca for their kind 
instruction how to use NSFCAM prior to actual observing runs, and 
to Dr. R. Nakamura for useful discussion about dust coagulation processes.
Drs. T. Nakajima, C. C. Dudley, and the anonymous referee gave useful 
comments on this manuscript.
MI was financially supported by the Japan Society for the Promotion 
of Science for his stays at the University of Hawaii.
Drs. A. T. Tokunaga and H. Ando gave MI the opportunity to work at 
the University of Hawaii.
This research has made use of the NASA/IPAC Extragalactic Database (NED) 
which is operated by the Jet Propulsion Laboratory, California Institute 
of Technology, under contract with the National Aeronautics and Space 
Administration.

\clearpage

%%%%%%%%%%%%%%%%%%%%%%        Tables       %%%%%%%%%%%%%%%%%%%%%%

\scriptsize

\begin{center}
\begin{deluxetable}{ccccccc}
\tablewidth{6.5in}
\tablenum{1}
\tablecaption{Observing Log}
\tablecolumns{7}
\tablehead{
\colhead{}& \colhead{}& \colhead{}& \multicolumn{2}{c}{Integration Time} 
& \multicolumn{2}{c}{Observing Date} 
\nl 
\colhead{}& \colhead{}& \colhead{$L$[OIII]} & \multicolumn{2}{c}{(sec)} 
& \multicolumn{2}{c}{(UT)} \nl 
\cline{4-5} \cline{6-7} %\nl
\colhead{Object}& \colhead{redshift}& \colhead{(10$^{42}$ erg s$^{-1}$)} 
& \colhead{$L$} & \colhead{$M'$} & \colhead{$L$} & \colhead{$M'$} \nl
\colhead{(1)}& \colhead{(2)}& \colhead{(3)}  & \colhead{(4)} 
& \colhead{(5)} & \colhead{(6)} & \colhead{(7)} %\nl
}
\startdata
3C 63  & 0.175 & 2 $^{a}$ & 1000 & 1200 & 1999 Nov 29 & 1999 Nov 29\\ 
3C 171 & 0.238 & 5 $^{a}$ & 1800 & 3600 & 1999 Nov 28 & 1999 Nov 29 \\
3C 195 (0806$-$10) & 0.110 & 5 $^{b}$ & 900 & 1200 & 1999 Nov 28 
& 1999 Nov 29 \\
3C 234 & 0.184 & 21 $^{a}$ & 400 & 800 & 1999 Nov 29 & 1999 Nov 29 \\ 
       &       &    & 200 &  400 & 2000 May 16 & 2000 May 16 \\ 
3C 321 & 0.096 & 2 $^{a}$ & 400 & 1600 & 2000 May 15 & 2000 May 15 \\
       &       & & 400 & 1800 & 2000 May 16 & 2000 May 16 \\
3C 445 & 0.056 & 1 $^{a}$ & 200 & 400 & 2000 May 15 & 2000 May 15 \\ 
3C 456 & 0.233 & 6 $^{a}$ & 1200 & 2400 & 1999 Nov 28 & 1999 Nov 29 \\ 
Cygnus A & 0.056 & 1 $^{c}$ & 600 & 2400 & 2000 Apr 19 & 2000 Apr 19 \\ 
         &       & & 400 & 1800 & 2000 May 16 & 2000 May 16 \\ 
Mrk 231 & 0.042 & 1 $^{d}$ & 150 & 200 & 2000 Apr 18 & 2000 Apr 18 \\ 
PG 1534+580 (Mrk 290) & 0.032 & 1 $^{d}$ & 300 & 800 & 2000 Apr 19 
& 2000 Apr 19 \\
PKS 1345+12 & 0.122 & 2 $^{d}$ & 600 & 800 & 2000 Apr 19 & 
2000 Apr 19 \\ 
\enddata
\tablecomments{
Column (1): Object name. 
Column (2): Redshift.
Column (3): Optical [OIII] emission line luminosity in 10$^{42}$ 
            erg s$^{-1}$ and references.
$^{a}$: Jackson \& Rawlings 1997; $^{b}$: Tadhunter et al. 1998; 
$^{c}$: Osterbrock \& Miller 1975; $^{d}$: Xu, Livio, \& Baum 1999.
Columns (4) and (5): Net on source integration time for $L$- and $M'$-band 
photometry in seconds, respectively.
Columns (6) and (7): Observing date for $L$- and $M'$-band photometry in UT, 
respectively.
}
\end{deluxetable}
\end{center}

\clearpage

\scriptsize

\begin{center}
\begin{deluxetable}{cccccc}
\tablewidth{6.5in}
\tablenum{2}
\tablecaption{Photometric Results}
\tablecolumns{6}
\tablehead{
\colhead{Object}& \colhead{$L$}& \colhead{$M'$}  & \colhead{$L - M'$} 
& \colhead{Aperture} & \colhead{Type} \nl
\colhead{}& \colhead{}& \colhead{}  & \colhead{} 
& \colhead{arcsec (kpc)} & \colhead{} \nl
\colhead{(1)}& \colhead{(2)}& \colhead{(3)}  & \colhead{(4)} 
& \colhead{(5)} & \colhead{(6)}  %\nl
}
\startdata
3C  63 & 13.4$\pm$0.3 & $>$11.0 & $<$2.7 & 6 (16.6) & Sy2 $^{a}$ \\ 
3C 171 & $>$13.9      & $>$10.9     &  \nodata & 6 (21.1) & Sy2 $^{a}$ \\
3C 195 & 10.7$\pm$0.1 & 9.9$\pm$0.1 & 0.8$\pm$0.1 & 6 (11.2) 
& Sy2 $^{b}$ \\
3C 234 & 10.7$\pm$0.1 & 9.5$\pm$0.2 & 1.2$\pm$0.2 & 6 (17.3) & Sy2 $^{c}$ \\
       & 10.6$\pm$0.1 & 9.3$\pm$0.2 & 1.3$\pm$0.2 & 6 (17.3) &\nodata \\
3C 321 $^{*}$ & 12.5$\pm$0.1 & 11.7$\pm$0.3 & 0.8$\pm$0.3 & 6 (10.0) &  Sy2 $^{a}$ \\ 
    & 12.5$\pm$0.1 & 11.4$\pm$0.2 & 1.1$\pm$0.2 & 6 (10.0) & \nodata \\ 
3C 445 & 9.2$\pm$0.1 & 8.4$\pm$0.1 & 0.8$\pm$0.1 & 6 (6.1) & Sy1 $^{a}$ \\ 
3C 456 & 12.7$\pm$0.2 & $>$10.1 & $<$2.8 & 6 (20.8) & Sy2 $^{a}$ \\ 
Cygnus A  & 12.7$\pm$0.1 & 11.4$\pm$0.3 & 1.3$\pm$0.3 & 6 (6.1) & Sy2 $^{d}$  \\
          & 12.0$\pm$0.1 & 11.1$\pm$0.2 & 0.9$\pm$0.2 & 6 (6.1) & \nodata \\
Mrk 231 & 7.4$\pm$0.1 & 6.4$\pm$0.1 & 1.0$\pm$0.1 & 6 (4.7) & Sy1 $^{e}$ \\ 
PG 1534+580 & 10.9$\pm$0.1 & 10.3$\pm$0.3 & 0.6$\pm$0.3 & 6 (3.6) & Sy1 $^{f}$ \\ 
PKS 1345+12 $^{*}$ & 11.5$\pm$0.1 & 10.7$\pm$0.2 & 0.8$\pm$0.2 & 6 (12.3) & Sy2 $^{e}$ \\ 
\enddata
\tablecomments{
Column (1): Object name. 
Column (2): $L$-band magnitude. 
            When sources are observed twice, both photometric results 
            are shown as independent data instead of combining these 
            data, to see time variation.
Column (3): $M'$-band magnitude.
Column (4): $L - M'$ color in magnitude.
Column (5): Aperture size in arcsec used for the photometry.
            Corresponding physical size in kpc is also shown 
            in parentheses.
Column (6): Optical spectral type and references.
Sy1 : type-1 AGNs. Sy2 : type-2 AGNs.
$^{a}$: Jackson \& Rawlings 1997; $^{b}$: Tadhunter et al. 1998; 
$^{c}$: Young et al. 1998; 
$^{d}$: Osterbrock 1983; $^{e}$: Veilleux, Kim, \& Sanders 1999a; 
$^{f}$: Xu, Livio, \& Baum 1999.   
Although 3C 234 shows broad optical emission lines, we classified this 
source as Sy2 following the reference. \\
\\
$^{*}$: 3C 321 and PKS 1345+12 have double nuclei with separations of 
a few arcsec (\cite{hec86,sco00}), but the $L$- and
$M'$-band fluxes of each object come predominantly from the main
nucleus.
}
\end{deluxetable}
\end{center}

\clearpage
\noindent 

\clearpage

%\small
\scriptsize

\begin{center}
\begin{deluxetable}{cccccccc}
\tablewidth{6.5in}
\tablenum{3}
\tablecaption{3--5 $\mu$m Photometric Results in the Literature.}
%\tablecolumns{60}
\tablecolumns{8}
\tablehead{
\colhead{Type}&\colhead{Object}& \colhead{Redshift} & \colhead{$L$} 
& \colhead{$M'$} & \colhead{$L - M'$} & \colhead{Aperture} 
& \colhead{Reference} \nl
\colhead{}&\colhead{}& \colhead{}& \colhead{} & \colhead{} & 
\colhead{} & \colhead{arcsec (kpc)} & \colhead{} \nl
\colhead{(1)}& \colhead{(2)}& \colhead{(3)}  & 
\colhead{(4)} & \colhead{(5)} & \colhead{(6)} & \colhead{(7)} 
& \colhead{(8)}  %\nl
}
\startdata
% Sy 1
Sy 1 & NGC 863 (Mrk 590) & 0.026 & 9.3 & 8.5 & 0.8 & 8, 5 (3.9, 2.5) & 1 \\
& NGC 931 (Mrk 1040) & 0.017 & 8.4 & 7.6 & 0.8 & 8, 5 (2.6, 1.6) & 1 \\
&          & & 8.9$\pm$0.1 & 7.9$\pm$0.2 & 1.0$\pm$0.2 & 7.9 (2.6) & 2 \\
& NGC 1365 & 0.005 & 7.6$\pm$0.1 & 6.9$\pm$0.1 & 0.7$\pm$0.1 & 9.1 (0.9) 
& 2 \\
&          & & 7.6$\pm$0.1 & 7.0$\pm$0.1 & 0.6$\pm$0.1 & 5 (0.5) & 3 \\
& NGC 3227 & 0.004 & 8.9$\pm$0.1 & 8.4$\pm$0.5 & 0.5$\pm$0.5 & 4.6 (0.4) 
& 2 \\
&          & & 8.6$\pm$0.1 & 8.0         & 0.6 & 15 (1.2) & 4 \\  
& NGC 3516 & 0.009 & 8.4$\pm$0.1 & 8.1$\pm$0.6 & 0.3$\pm$0.6 & 
15, 10 (2.6, 1.7)& 4 \\
& NGC 4051 & 0.002 & 9.0$\pm$0.1 & 8.0 & 1.0 & 15 (0.6) & 4\\
& NGC 4151 & 0.003 & 7.4$\pm$0.1 & 6.4$\pm$0.1 & 1.0$\pm$0.1 & 7.9 (0.5) 
& 2 \\
&          & & 7.3$\pm$0.1 & 6.4$\pm$0.1 & 0.9$\pm$0.1 & 10 (0.6) & 4 \\  
& NGC 5548 & 0.017 & 8.6$\pm$0.1 & 8.0$\pm$0.3 & 0.6$\pm$0.3 & 7.9 (2.6) & 2 \\
&          & & 9.0$\pm$0.1 & 8.0$\pm$0.7 & 1.0$\pm$0.7 & 10 (3.2) & 4 \\
& NGC 6814 & 0.005 & 9.8$\pm$0.1 & 8.9$\pm$0.6 & 0.9$\pm$0.6 & 7.9 (0.8) & 2 \\
& NGC 7469 & 0.016 & 8.1$\pm$0.1 & 7.0$\pm$0.2 & 1.1$\pm$0.2 & 7.9 (2.4) & 2 \\
&          & & 8.0$\pm$0.1 & 7.4$\pm$0.1 & 0.6$\pm$0.1 & 5 (1.5) & 3 \\
& 3A 0557--385 & 0.034 & 8.3    & 7.1         & 1.2 & 5 (3.2) & 1 \\
& 3C 120 & 0.033 & 9.2$\pm$0.1   & 8.4$\pm$0.7 & 0.8$\pm$0.7 & 15 (9.2) & 4 \\
& 3C 273 & 0.158 & 8.0$\pm$0.1   & 7.3$\pm$0.2 & 0.7$\pm$0.2 & 9.1 (23.2)& 2 \\
& 3C 445 & 0.056 & 9.5           & 8.6         & 0.9   & 8, 5 (8.1, 5.1) 
& 5 \\
& Akn 120 & 0.032 & 9.0$\pm$0.1  & 8.1$\pm$0.3 & 0.9$\pm$0.3 & 4.6 (2.8)& 2 \\
& ESO 113--IG45 & 0.045 & 8.4$\pm$0.1 & 8.1$\pm$0.3 & 0.3$\pm$0.3 & 
9.1 (7.5) & 2 \\
& IC 4329A & 0.016 & 7.8$\pm$0.1 & 7.5$\pm$0.1 & 0.3$\pm$0.1 & 
4.6 (1.4) & 2 \\
&         & & 7.9$\pm$0.1 & 6.9$\pm$0.3 & 1.0$\pm$0.3 & 15 (4.6) & 4 \\
& MCG$-$2--58--22 & 0.048 & 9.3 & 8.6 & 0.7 & 8, 5 (7.0, 4.4) & 1 \\ 
& MCG 8--11--11 & 0.020 & 8.9 & 7.9 & 1.0 & 8, 5 (3.0, 1.9) & 1 \\
&      & & 8.8$\pm$0.1 & 8.1$\pm$0.3 & 0.7$\pm$0.3 & 7.9 (3.0) & 2 \\
&      & & 8.8$\pm$0.1 & 7.3$\pm$0.2 & 1.5$\pm$0.2 & 15 (5.7) & 4 \\
& Mrk 79 & 0.022 & 9.3 & 8.4 & 0.9 & 8 (3.3) & 1 \\
& Mrk 231 & 0.042 & 7.4$\pm$0.1 & 6.5$\pm$0.1 & 0.9$\pm$0.1 & 10 (7.8) & 4 \\
& Mrk 335 & 0.026 & 8.7$\pm$0.1 & 7.6$\pm$0.2 & 1.1$\pm$0.2 & 7.9 (3.9) & 2 \\
& Mrk 359  & 0.017 & 10.3$\pm$0.1 & 9.2$\pm$0.1 & 1.1$\pm$0.1 & 5 (1.6) & 6 \\
& Mrk 1152 & 0.053 & 10.6 & 9.4 & 1.2 & 8, 5 (7.7, 4.8) & 1 \\
\hline 
% Sy 2 
Sy 2 & NGC 262 (Mrk 348) & 0.015 & 10.5$\pm$0.1 & 9.1$\pm$0.1 & 1.4$\pm$0.1 & 3 (0.9) & 7 \\
& NGC 526a & 0.019 & 9.6  & 8.4 & 1.2 & 8, 5 (2.9, 1.8) & 1 \\
& NGC 1052 & 0.005 & 9.8$\pm$0.1  & 9.1$\pm$0.2  & 0.7$\pm$0.2  & 3 (0.3)   & 7 \\
& & & 9.7$\pm$0.1 & 9.3$\pm$0.2 & 0.4$\pm$0.2 & 4 (0.4) & 8 \\
\tablebreak
& NGC 1068 & 0.004 & 5.3 & 3.7 & 1.6$\pm$0.4  & 0.6 (0.1) & 9 \\
&          & & 4.5$\pm$0.1  & 3.2$\pm$0.1  & 1.3$\pm$0.1  & 3 (0.2) & 7 \\
& NGC 1275 & 0.018 & 8.1$\pm$0.1 & 7.1$\pm$0.1 & 1.0$\pm$0.1 & 7.9 (2.7)  & 2 \\
& NGC 1808 & 0.003 & 8.7$\pm$0.1 & 8.7$\pm$0.2 & 0.0$\pm$0.2 & 5 (0.3) & 3 \\
& NGC 2992 & 0.008 & 9.2 & 8.4 & 0.8 & 6 (0.9) & 1 \\
&          & & 10.0$\pm$0.1 & 9.2$\pm$0.3  & 0.8$\pm$0.3  & 3 (0.5) &  7 \\
& NGC 3094 & 0.008 & 8.2$\pm$0.1  & 7.5$\pm$0.1  & 0.7$\pm$0.1  & 5 (0.8) & 3 \\
& NGC 3281 & 0.011 & 8.4          & 7.2          & 1.2 & 6 (1.3) & 10 \\
& NGC 4418 $^{a}$ & 0.007 & 11.1$\pm$0.1 & 10.2$\pm$0.2 & 0.9$\pm$0.2  
& 5 (0.7) & 3 \\
& NGC 4736 & 0.001 & 7.0$\pm$0.1  & 7.4$\pm$0.4 & $-$0.4$\pm$0.4 
& 15 (0.3)  & 4 \\
& NGC 4945 & 0.002 & 8.3 & 7.5 & 0.8 & 7.5 (0.3) & 11 \\
& NGC 4968 & 0.010 & 10.0$\pm$0.1 & 8.7$\pm$0.2 & 1.3$\pm$0.2 & 3 (0.6) & 7 \\
& NGC 5252 & 0.023 & 10.6$\pm$0.1 & 10.0$\pm$0.3 & 0.6$\pm$0.3 & 3 (1.3) & 7 \\
& NGC 5506 & 0.006 & 7.6$\pm$0.1  & 6.7$\pm$0.1 & 0.9$\pm$0.1 & 4.6 (0.5)& 2 \\
&          & & 7.1$\pm$0.1  & 6.2$\pm$0.1 & 0.9$\pm$0.1 & 3 (0.4) & 7  \\
& NGC 7130 (IC 5135) & 0.016 & 10.0$\pm$0.1 & 9.9$\pm$0.2 & 0.1$\pm$0.2 & 7.8 (2.4) & 3 \\
& NGC 7172 & 0.009 & 9.1$\pm$0.1  & 8.5$\pm$0.1 & 0.6$\pm$0.1 & 5 (0.9) & 3 \\
&          & & 8.2          & 7.3$\pm$0.1 & 0.9$\pm$0.1 & 8, 5 (1.4, 0.9) & 6 \\
&          & & 9.5$\pm$0.1  & 8.6$\pm$0.1 & 0.9$\pm$0.1  & 3 (0.5) & 7 \\
& NGC 7314 & 0.005 & 10.3 & 9.4 & 0.9 & 5 (0.5) & 1 \\
& NGC 7479 & 0.008 & 9.8$\pm$0.1  & 8.6$\pm$0.1 & 1.2$\pm$0.1 & 5 (0.8) 
& 3 \\
& NGC 7582 & 0.005 & 7.8$\pm$0.1  & 7.1$\pm$0.1 & 0.7$\pm$0.1 & 9.1 (0.9) & 2 \\
& NGC 7674 (Mrk 533) & 0.029 & 9.0$\pm$0.1 & 8.2$\pm$0.1 & 0.8$\pm$0.1 
& 5 (2.7) & 3 \\ 
&  & & 9.1$\pm$0.1  & 8.0$\pm$0.1 & 1.1$\pm$0.1 & 3 (1.6) & 7 \\
& 3C 33    & 0.060 & 11.9$\pm$0.1 & 11.2$\pm$0.1 & 0.7$\pm$0.1   & 3 (3.3) & 12 \\ 
& 3C 223   & 0.137 & 12.9$\pm$0.1 & 11.8$\pm$0.2 & 1.1$\pm$0.2   & 3 (6.8) & 12 \\ 
& 3C 234 & 0.185 & 10.9         & 9.3         & 1.6 & 6 (17.4) & 5  \\
&        & & 10.5$\pm$0.1 & 9.9$\pm$0.1 & 0.6$\pm$0.1 & 3 (8.7) & 12 \\ 
& Circinus & 0.002 & 6.4 & 5.1 & 1.3 & 5 (0.2) & 11 \\
& IRAS 00198$-$7926 & 0.073 & 9.8 & 7.7 & 2.1 & 4.7 (6.1) & 13 \\
& IRAS 00521$-$7054 & 0.069 & 9.2 & 8.1 & 1.1 & 6.2 (7.6) & 13 \\
& IRAS 04385$-$0828 & 0.015 & 9.5 & 8.0 & 1.5 & 4.7 (1.3) & 13 \\
& IRAS 05189$-$2524 & 0.042 & 8.1$\pm$0.1 & 7.4$\pm$0.1 & 0.7$\pm$0.1 
& 5 (3.9) & 3 \\ 
&             & & 8.5 & 7.2 & 1.3 & 4.7 (3.6) & 13 \\
& IRAS 08572+3915 $^{b}$ & 0.058 & 9.2$\pm$0.3 & 8.0$\pm$0.3 & 1.2$\pm$0.4 & 4.5 (4.7) & 14 \\ 
& IRAS 20460+1925 & 0.181 & 9.2 & 8.3 & 0.9 & 11.9 (33.9) & 13 \\  
& MCG$-$5--23--16 & 0.008 & 8.5$\pm$0.1 & 7.7$\pm$0.1 & 0.8$\pm$0.1 
& 3 (0.5) & 7 \\
& Mrk 573 & 0.017 & 10.1$\pm$0.1 & 9.0$\pm$0.3 & 1.1$\pm$0.3 
& 3 (1.0) & 7 \\ 
\hline
\tablebreak
% SB, LINER, composite
SB       & NGC 520 & 0.008 & 9.6$\pm$0.1 & 9.1$\pm$0.2 & 0.5$\pm$0.2 
& 7.8 (1.2) & 3 \\
LINER    & NGC 613 & 0.005 & 9.5$\pm$0.1 & 8.7$\pm$0.1 & 0.8$\pm$0.1 
& 5 (0.5) & 3 \\
Unknown  & NGC 660 & 0.003 & 8.9$\pm$0.1    & 8.7$\pm$0.1 & 0.2$\pm$0.1 
& 5 (0.3) & 3 \\
& NGC 828  & 0.018 & 10.4$\pm$0.1   & 10.6$\pm$0.1 & $-$0.2$\pm$0.1 
& 5 (1.7) & 3 \\
& NGC 1614 (Mrk 617) & 0.016 & 9.0$\pm$0.1 & 8.6$\pm$0.1 & 0.4$\pm$0.1 
& 7.8 (2.4) & 3 \\
&     &   & 9.2 & 8.8$\pm$0.1 & 0.4$\pm$0.1  & 5 (1.5) & 6 \\
& NGC 2110  & 0.008 & 9.4 & 8.9$\pm$0.2 & 0.5$\pm$0.2 & 6 (0.9) & 6 \\
& NGC 2339  & 0.007 & 9.7$\pm$0.1  &  9.5$\pm$0.2 & 0.2$\pm$0.2 & 7.8 (1.1) & 3 \\
& NGC 2388  & 0.014 & 10.1$\pm$0.1 &  9.8$\pm$0.2 & 0.3$\pm$0.2 & 5 (1.3) & 3 \\
& NGC 2623  & 0.018 & 10.6$\pm$0.1 & 10.2$\pm$0.2 & 0.4$\pm$0.2 & 5 (1.7) & 3 \\
& NGC 2782  & 0.009 & 10.1$\pm$0.1 &  9.6$\pm$0.1 & 0.5$\pm$0.1  
& 5 (0.9)& 3 \\
& NGC 3079  & 0.004 & 9.2          &  9.4$\pm$0.2 & $-$0.2$\pm$0.2 
& 6 (0.5) & 6 \\
& NGC 4102 & 0.003 & 8.5$\pm$0.1 & 8.3$\pm$0.1 & 0.2$\pm$0.1 & 5 (0.3) & 3 \\
& NGC 4194 (Mrk 201) & 0.008 & 9.4$\pm$0.1 & 8.9$\pm$0.1 & 0.5$\pm$0.1 
& 5 (0.8) & 3 \\ 
& NGC 4579 & 0.005 & 9.4 & 9.5$\pm$0.3 & $-$0.1$\pm$0.3 & 6 (0.6) & 6 \\
& NGC 4826 & 0.001 & 8.9 & 9.2$\pm$0.2 & $-$0.3$\pm$0.2 & 6 (0.1) & 6 \\
& NGC 6764 & 0.008 & 10.9 & 9.9$\pm$0.2 & 1.0$\pm$0.2 & 5 (0.8) & 6 \\ 
& NGC 7714 & 0.009 & 10.4$\pm$0.1 & 10.1$\pm$0.2 & 0.3$\pm$0.2 & 5 (0.9) 
& 3 \\ 
& NGC 7770 & 0.014 & 11.2$\pm$0.1   & 10.5$\pm$0.2 & 0.7$\pm$0.2 
& 5 (1.3) & 3 \\ 
& NGC 7771 & 0.014 & 10.3$\pm$0.1   & 9.9$\pm$0.1  & 0.4$\pm$0.1 
& 5 (1.3) & 3 \\
& MCG$-$3--4--14 & 0.033 & 10.7$\pm$0.1 & 10.2$\pm$0.2 & 0.5$\pm$0.2 
& 5 (3.1) & 3 \\ 
& Mrk 331 & 0.018 & 9.8$\pm$0.1    & 9.5$\pm$0.2  & 0.3$\pm$0.2 
& 5 (1.7) & 3 \\ 
& UGC 3094 & 0.025 & 10.6$\pm$0.1   & 10.2$\pm$0.2 & 0.4$\pm$0.2 
& 5 (2.4)& 3 \\
\enddata
\tablecomments{
Column (1): Spectral type. Sy1: type-1 AGNs.
Sy2: Dust-obscured AGNs either based on optical spectral type 
or detailed studies at other wavelengths.
SB: starburst galaxies.
LINER: LINER-type galaxies.
Column (2): Object name. 
Column (3): Redshift. 
Column (4): $L$- or $L'$-band magnitude. 
Column (5): $M'$- or $M$-band magnitude. 
Column (6): $L - M'$ color in magnitude. 
Column (7): Aperture size in arcsec and corresponding physical size in 
            kpc (in parentheses) used for the photometry. 
Column (8): Reference for the photometry.
1: Ward et al. 1987; 2: McAlary, McLaren, \& McGonegal 1983; 
3: Dudley 1998; 4: McAlary, McLaren, \& Crabtree 1979; 
5: Elvis et al. 1984; 6: Lawrence et al. 1985; 
7: Alonso-Herrero et al. 2000; 
8: Becklin, Tokunaga, \& Wynn-Williams 1982; 
9: Marco \& Alloin 2000; 10: Simpson 1998a; 
11: Moorwood \& Glass 1984; 12: Simpson et al. 2000; 
13: Vader et al. 1993; 14: Dudley \& Wynn-Williams 1997. \\
%\vspace{1cm}
\\ 
$^{a}$: Although NGC 4418 is optically a normal star-forming galaxy 
        (Rush, Malkan, \& Spinoglio 1993), 
        we classify this source as a Sy 2 based on a detailed infrared 
        study by Dudley \& Wynn-Williams (1997). \\
$^{b}$: Although IRAS 08572+3915 is optically a LINER 
        (Veilleux et al. 1995), 
        we classify this source as a Sy 2 based on detailed infrared 
        studies by Dudley \& Wynn-Williams (1997) and 
        Imanishi \& Ueno (2000).
}
\end{deluxetable}
\end{center} 

%%%%%%%%%%%%%%%%%%%%%%%%%% Figures %%%%%%%%%%%%%%%%%%%%%%%%%%%%%%%%

\begin{figure*}
\centerline{\psfig{file=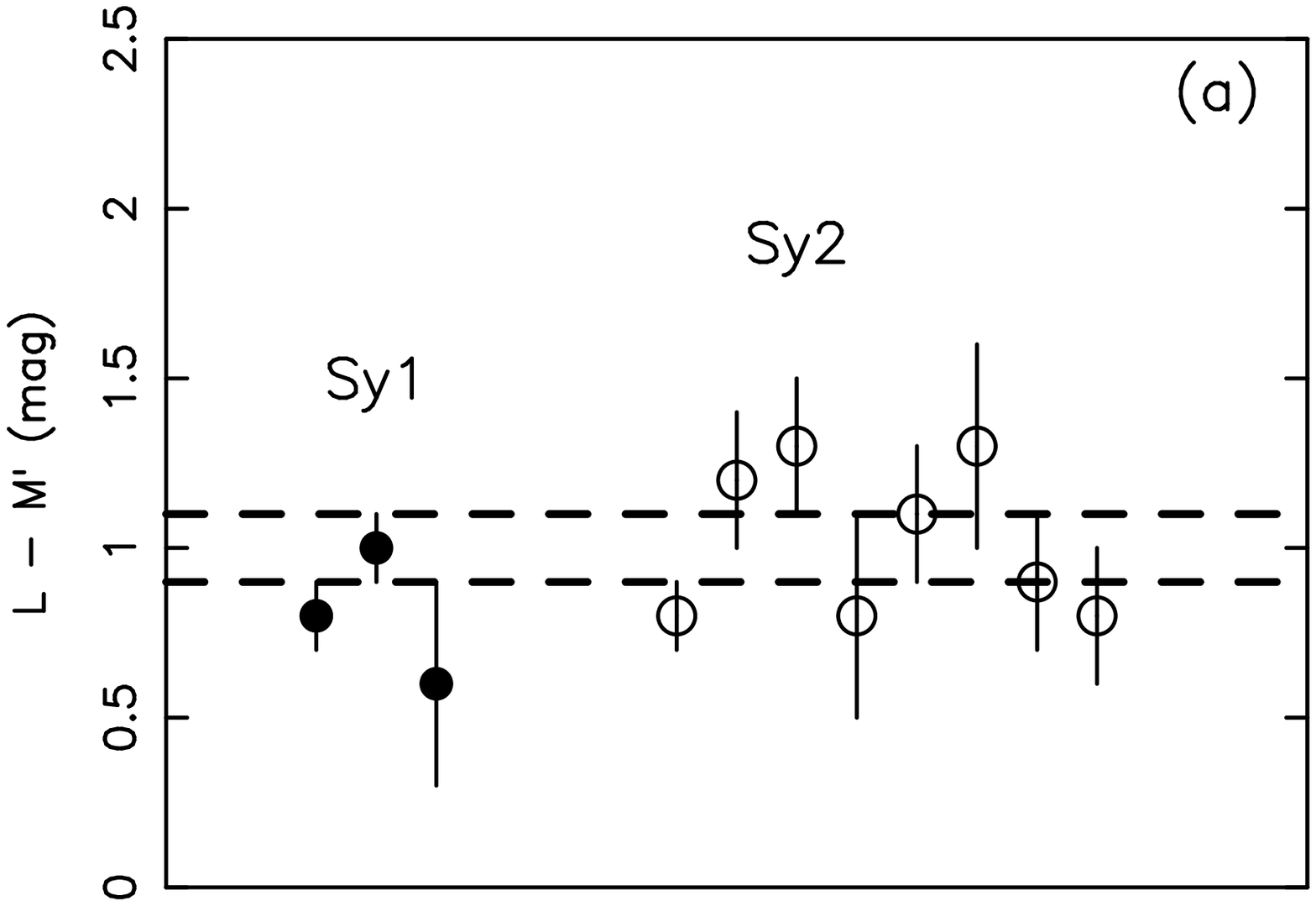,angle=0,width=4.2in}}
\centerline{\psfig{file=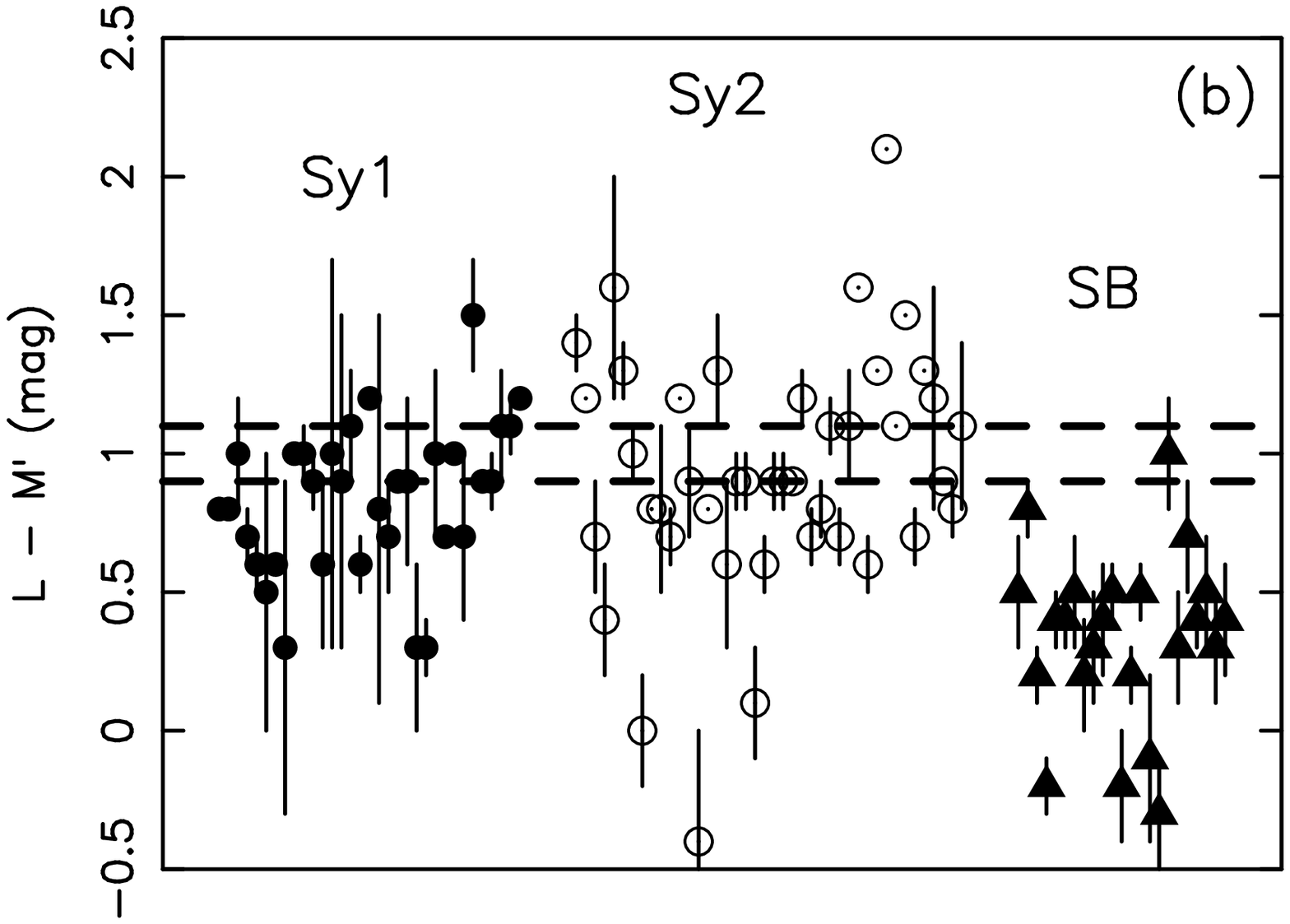,angle=0,width=4.2in}}
\caption{
(a): Distribution of $L - M'$ colors of type-1 and type-2 AGNs 
that we observed (Table 2). 
Filled circles: type-1 AGNs.
Open circles: type-2 AGNs. 
The region between the two dashed lines denotes the intrinsic $L - M'$ color 
(= 1.0$\pm$0.1) of AGNs in the case of no dust extinction 
(see $\S$ 4). 
3C 234, 3C 321, and Cygnus A were observed twice, and these two data points
are plotted independently.
3C 63, 3C 171, and 3C 456 are excluded because their $L - M'$ colors 
were not derived directly from the observed data.
(b) $L - M'$ colors of AGNs and starburst/LINER galaxies taken from 
the literature (Table 3).
Filled circles: type-1 AGNs.
Open circles: type-2 AGNs.
Filled triangles: starbursts, LINERs and galaxies with unknown spectral 
type.
The region between the two dashed lines denotes the intrinsic $L - M'$ 
(= 1.0$\pm$0.1) color of AGNs in the case of no dust extinction 
(see $\S$ 4). 
}
\end{figure*} 

\begin{figure*}
\centerline{\psfig{file=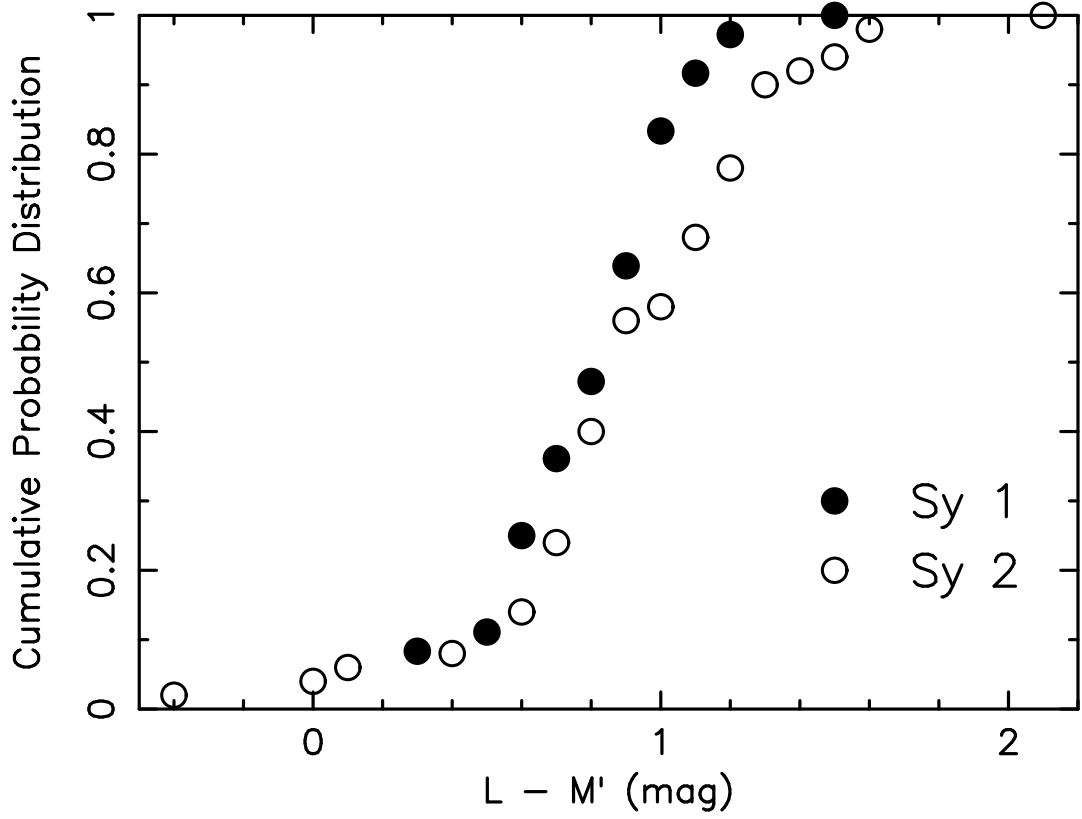,angle=0,width=6.5in}}
\caption{
Cumulative probability distribution of $L - M'$  colors.
Filled circles: type-1 AGNs (36 data points).
Open circles: type-2 AGNs (50 data points).
}
\end{figure*} 

\clearpage

\begin{figure*}
\centerline{\psfig{file=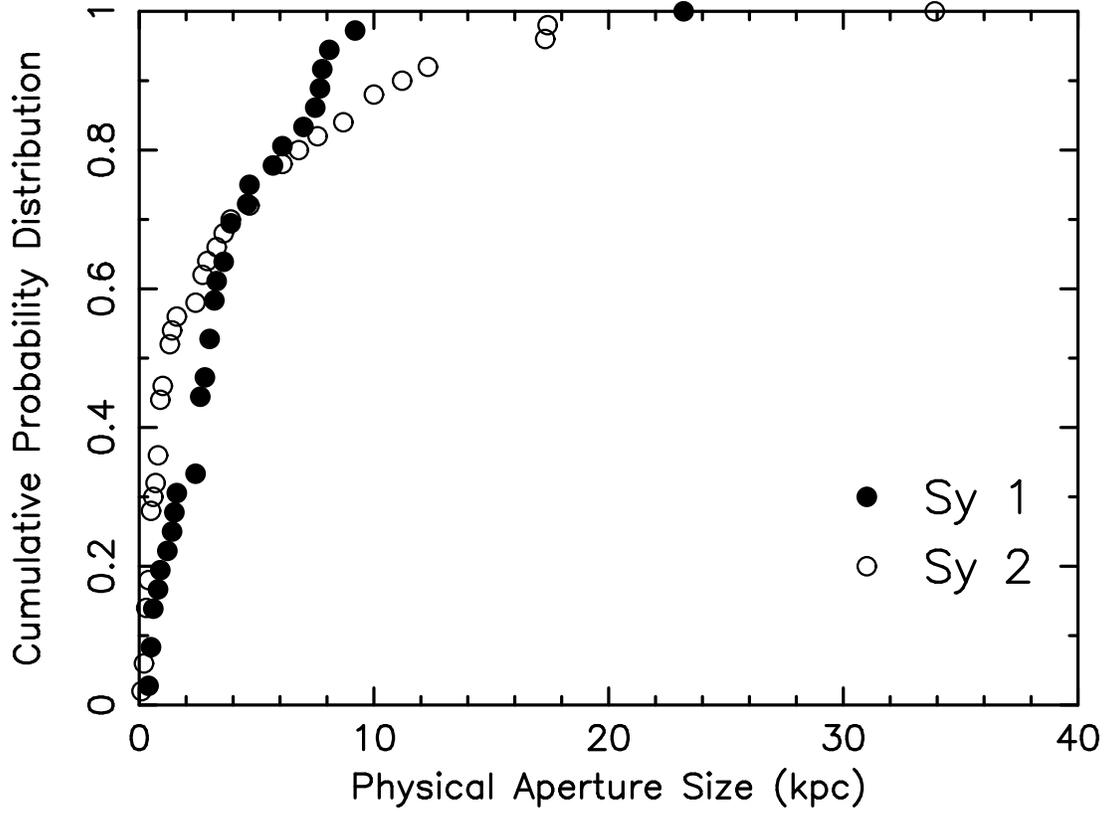,angle=0,width=6.5in}}
\caption{
Cumulative probability distribution of physical aperture size in kpc.
Filled circles: type-1 AGNs.
Open circles: type-2 AGNs.
}
\end{figure*} 

\clearpage

\begin{figure*}
\centerline{\psfig{file=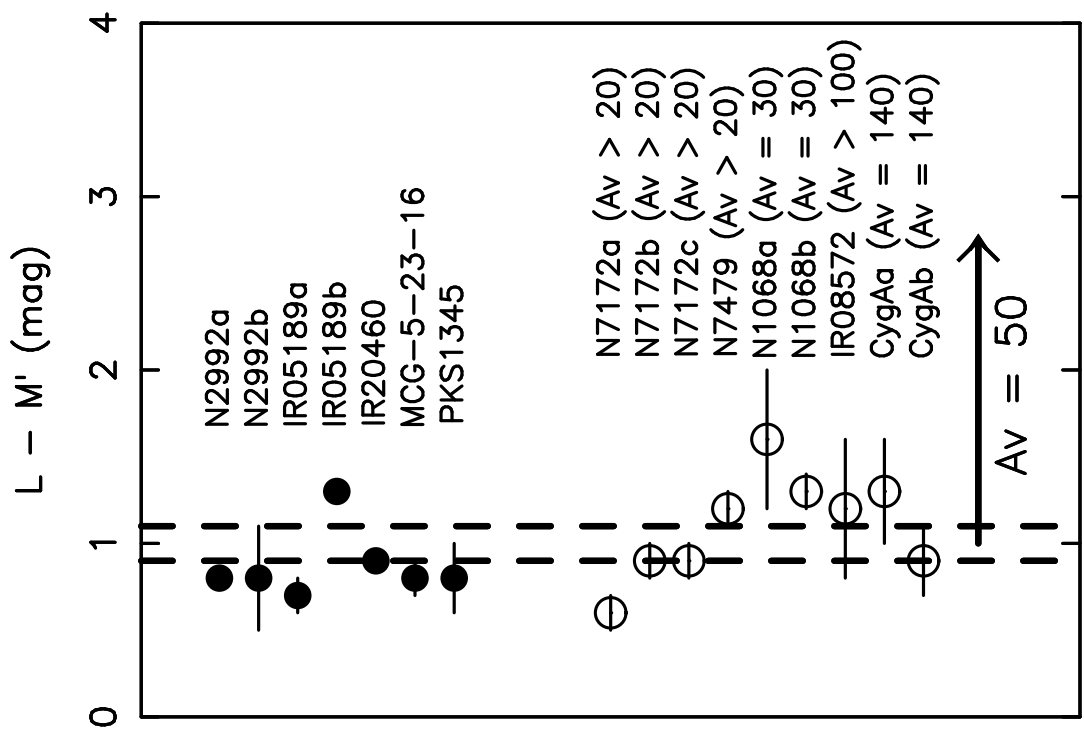,angle=0,width=6.5in}}
\caption{
$L - M'$ colors of the representative samples of type-2 AGNs.
Filled circles: less obscured type-2 AGNs for which broad Pa$\alpha$ 
or Pa$\beta$ emission components are detected at $<$2 $\mu$m. 
Open circles: highly dust-obscured type-2 AGNs.
Estimated $A_{\rm V}$ values are shown in the figure.
See the text ($\S$ 5.1) for more details.
The length of the upper arrow corresponds to the $L - M'$ color excess 
by dust extinction with $A_{\rm V}$ = 50 mag for the Galactic dust 
extinction curve.
}
\end{figure*} 

\end{document}